\documentclass[twocolumn,english,aps,prb,floats]{revtex4-2}
\usepackage[T1]{fontenc}
\usepackage[latin9]{inputenc}
\usepackage{babel}
\usepackage{textcomp}
\usepackage{amsmath}
\usepackage{amssymb}
\usepackage{graphicx}
\usepackage{wasysym}
\usepackage{esint}
\usepackage[unicode=true,
 bookmarks=false,
 breaklinks=false,pdfborder={0 0 1},backref=section,colorlinks=false]
 {hyperref}

\makeatletter

\newcommand{\lyxmathsym}[1]{\ifmmode\begingroup\def\b@ld{bold}
  \text{\ifx\math@version\b@ld\bfseries\fi#1}\endgroup\else#1\fi}

\usepackage{babel}
\DeclareMathOperator{\sign}{sign}
\usepackage{siunitx}
\DeclareSIUnit\angstrom{\text{\AA}}
\usepackage{xcolor}

\@ifundefined{textcolor}{}{%
 \definecolor{BLACK}{gray}{0}
 \definecolor{WHITE}{gray}{1}
 \definecolor{RED}{rgb}{1,0,0}
 \definecolor{GREEN}{rgb}{0,1,0}
 \definecolor{BLUE}{rgb}{0,0,1}
 \definecolor{CYAN}{cmyk}{1,0,0,0}
 \definecolor{MAGENTA}{cmyk}{0,1,0,0}
 \definecolor{YELLOW}{cmyk}{0,0,1,0}
}

\usepackage{ifpdf}\usepackage{bm}

\makeatother

\begin{document}
\title{Giant Barnett Effect from Moving Dislocations}
\author{Eugene M. Chudnovsky and Jorge F. Soriano}
\affiliation{Physics Department, Herbert H. Lehman College and Graduate School,
The City University of New York, 250 Bedford Park Boulevard West,
Bronx, New York 10468-1589, USA }
\date{\today}
\begin{abstract}
We show that moving dislocations generate giant effective local magnetic fields in a crystal lattice that can flip spins. Since massive creation of fast-moving dislocations is associated with a powerful elastic stress, this suggests a new mechanism of the magnetization reversal generated by laser or microwave beams or by electrically induced shear deformation. 
\end{abstract}
\maketitle

\section{Introduction}\label{intro}

\begin{figure}\centering
\includegraphics[width=0.8\linewidth]{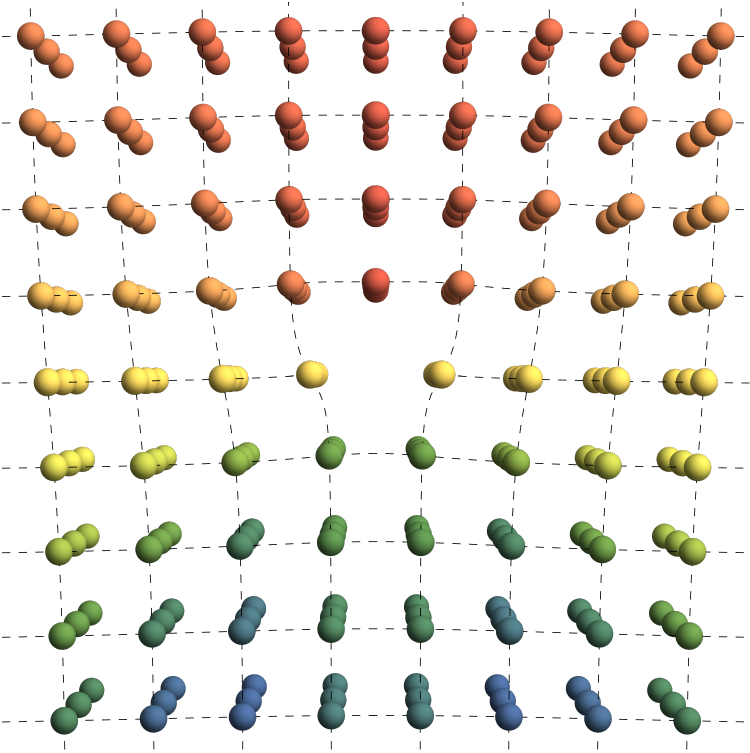}
\caption{}
\label{edgeDislocation}
\end{figure}
There exists a large body of research on switching spins in solids by short electric pulses \cite{Yang-Science2017}, laser beams \cite{Xu-JMMM2022}, microwaves  \cite{Cai-PRB2013,Miyashita-PRL2023}, surface acoustic waves \cite{Tejada-EPL2017,Camara-PRA2019}, and chiral phonons. The latter comes as no surprise since phonons can carry angular momentum \cite{Zhang-PRL2014,DG-EC-PRB2015,Nakane-PRB2018,DG-EC-PRB2021} which they can transfer to the spins. While in some cases a reasonable understanding of underlying mechanisms was achieved, in other cases they remained obscure. 

Recently, Davies et al. \cite{Davies-Nature2024} have elucidated the importance of the Barnett effect \cite{Barnett} in phononic switching of spins in a paramagnetic substrate. They suggested that circularly polarized phonons could generate local elastic twists that in the local coordinate frame of the crystal lattice are equivalent to the magnetic fields capable of flipping the spins. Large effective magnetic fields from chiral phonons were previously observed in $4f$ paramagnets \cite{Juraschek-PRR2022} and rare-earth halides \cite{Luo-Science2023}. 

Implications of the Barnett effect and the reciprocal Einstein--de Haas effect \cite{Einstein1,Einstein2} at the nanoscale have been discussed in various contexts in the past. They have been shown to provide a universal mechanism of the interaction of chiral phonons with spins \cite{EC-PRB2002,EC-PRL2004,EC-DG-PRL2004,EC-DG-RS-PRB2005,Dornes-Nature2019,Tauchet-Nature2022} and helped understand the dynamics of angular momentum in nanocantilevers \cite{Wallis-APL2006,JCG-PRB2009,Mori-PRB2020} and molecular magnets, including experiments with a single magnetic molecule attached to a suspended carbon nanotube \cite{Wernsdorfer-2015}. Barnett effect in thin magnetic films and nanostructures has been studied by Bretzel et al. \cite{Bretzel-APL2009}. They observed that the rotational frequencies required to switch magnetization in conventional materials would be beyond present experimental possibilities. 

In this paper we show that a moving edge dislocation, can generate an effective magnetic field in excess of $10\,\mathrm T$ in the coordinate frame of the crystal lattice. This can be easily understood from the following argument. The angle (in radians) by which the crystallographic axes are rotated near the core of a static edge dislocation shown in Fig.\ \ref{edgeDislocation} is of the order of $\delta \phi \sim 1/(2\pi)$. When a moving dislocation passes at a speed $c$ through a unit cell of the crystal lattice of size $a$, the crystallographic axes rotate on a time scale $\delta t \sim a/c$. The angular velocity of that rotation, $\Omega$, is of order $\delta\phi/\delta t \sim c/(2\pi a)$. Since the spin states are formed by the interactions in the rotating coordinate frame of the crystal lattice, the effect of the rotation on the spin is equivalent \cite{hehlinertial1990,EC-DG-RS-PRB2005} to the action of the effective magnetic field $H = \Omega/\gamma \sim c/(2\pi a \gamma)$, where $\gamma$ is the gyromagnetic ratio. The direction of the spin then relaxes due to spin-lattice and other interactions toward the direction of the effective field formed by the rotation, the magnetic anisotropy of the crystal lattice, and the external magnetic field. Such relaxation in the non-inertial coordinate frame is the essence of the Barnett effect.  

Under a large elastic stress, dislocations can move at a speed comparable to the speed of sound \cite{Benat2021}. Taking $c \sim 3 \times 10^3\,\unit{\meter\per\second}$, $a \sim 3 \si{\angstrom}$, and $\gamma =1.6 \times 10^{11}\,\unit{\per\second\per\tesla}$,  one obtains $H \sim c/(2\pi a \gamma) \sim 10\,\mathrm T$, which must be sufficient to flip any spin subjected to such a field. As shown below, $H$ goes down as $(a/r)^2$ with the distance $r$ from the center of the dislocation, still producing large fields capable of flipping spins several lattice spacings away from the dislocation core. It has been observed that the shock stress from an ultrashort laser pulse produces a highly dense rapidly-developing dislocation structure \cite{Matsuda-JAP2014}. We believe that the dislocation-induced spin-flipping mechanism may have been overlooked in the previous studies of the magnetization reversal by laser pulses and other means of creation of the elastic stress. 

The paper is organized as follows. The effective magnetic fields generated in a solid by the elastic twist from a moving edge dislocation are computed in Section \ref{reversal}. The flipping of a spin by an edge dislocation passing nearby is demonstrated in Section \ref{reversal} via the solution of the Landau-Lifshitz equation. Implications of our results for experiments are discussed in Section \ref{discussion}. 

\section{Effective magnetic fields generated by a moving edge dislocation} \label{Barnett}
\subsection{Stationary dislocation}
We start from a periodic, non-dislocated cubic lattice of constant $a$, with $N$ atoms at positions $\mathbf x_i$, for $i=1,\dots,N$. In the presence of a dislocation, atoms move from $\mathbf x_i$ to $\tilde{\mathbf x}_i$. The displacement field $\mathbf u(\mathbf x)$ is such that $\tilde{\mathbf x}_i=\mathbf x_i+\mathbf u(\mathbf x_i)$.

We consider an edge dislocation, in which an extra half-plane of atoms is inserted in the lattice parallel to the $y-z$ plane for $x=0$ and $y>0$, as shown in Fig.~\ref{edgeDislocation}.

The equilibrium displacement field configuration for this type of dislocation was first derived by J.~M.~Burgers~\cite{burgerssome1939}, and later analyzed elsewhere (e.\,g.~\cite{landautheory1986}). Expressions therein differ only in their choice of boundary and continuity conditions. In this work, we chose conditions such that  \emph{(i)} the function is continuous everywhere except at the origin and the half plane specified by $x=0$ and $y>0$, and \emph{(ii)} the solution is symmetric under $x\to-x$ reflections, consistent with a dislocation placed at $x=0$. Under these conditions, and writing $\mathbf u=\left<u_x,u_y,u_z\right>$,  we adopt the expressions
\begin{subequations}
	\begin{gather}
		u_x=\frac{a}{2\pi}\left(\arctan_2(-y,x)+\frac{1}{2(1-\sigma)}\frac{xy}{x^2+y^2}\right),
		\label{eq:displacement-static-x-final}\\
		\begin{split}
		u_y=-\frac{a}{2\pi}\left(\frac{1-2\sigma}{2(1-\sigma)}\right.&\ln\sqrt{x^2+y^2}\\&+\left.\frac{1}{2(1-\sigma)}\frac{y^2}{x^2+y^2}\right),\end{split}\label{eq:displacement-static-y-final}
	\end{gather}\label{eq:displacement-static-final}%
\end{subequations}
and $u_z=0$. Here, $\sigma$ is the Poisson ratio, which may be expressed in terms of the Lam\'e coefficients $\lambda$ and $\mu$ as $\sigma=\lambda/2(\lambda+\mu)$. The function
\begin{equation}
	\arctan_2(x,y)\equiv
	\begin{cases}
		\arctan\frac yx,\ x\geq0,\\
		\arctan\frac yx\pm\sign y,\ x<0,
	\end{cases}
\end{equation}
where $\arctan z$ is the single-valued arctangent function with co-domain $[-\pi/2,\pi/2]$,
ensures continuity at $x=0$ for $y<0$, while keeping the dislocation-induced discontinuity for $y>0$.

\subsection{Moving dislocation}
\subsubsection{Displacement field}
The displacement field produced by an edge dislocation moving with constant speed $c$ along the $x$ axis may be easily obtained replacing $x$ by $x-c t$ in \eqref{eq:displacement-static-final}. Nevertheless, at high velocities, corrections must be included to account for the speed limits set by transversal and longitudinal wave propagation.
These where first worked out by J.~D.~Eshelby~\cite{eshelbyuniformly1949}. Different, compatible expressions were found elsewhere for different boundary conditions (e.\,g.~\cite{nabarrotheory1967}). Here, we adapt those to ensure consistency with our stationary solution \eqref{eq:displacement-static-final}:
\begin{subequations}
	\begin{gather}
		\begin{split}
			u_x=\frac{a}{\pi}\frac{c_t^2}{c^2}\left(\vphantom{\tilde\gamma_t^2}\right.\arctan_2&(-\gamma_l y,x-ct)\\&\left.-\tilde\gamma_t^2\arctan_2(-\gamma_t y,x-ct)\right),\end{split}\\
		\begin{split}
			u_y=\frac{a}{\pi}\frac{c_t^2}{c^2}\left(\vphantom{\frac{\tilde\gamma_t^2}{\gamma_t}}\right.\gamma_l\ln&\sqrt{(x-ct)^2+\gamma_l^2y^2}\\&-\left.\frac{\tilde\gamma_t^2}{\gamma_t}\ln\sqrt{(x-ct)^2+\gamma_t^2y^2}\right),
		\end{split}
	\end{gather}\label{eq:eshelby}%
\end{subequations}
where
\begin{equation}
	\gamma_t=\sqrt{1-\frac{c^2}{c_t^2}} ,			\quad
		\tilde\gamma_t=\sqrt{1-\frac{c^2}{2c_t^2}},	\quad
		\gamma_l=\sqrt{1-\frac{c^2}{c_l^2}},
\end{equation}
and $c_t$ and $c_l$ are the speed of transversal and longitudinal waves, respectively, given by
\begin{equation}
		c_t=\sqrt{\frac{\mu}{\rho}},\qquad
		c_l=\sqrt{\frac{\lambda+2\mu}{\rho}}
\end{equation}
The $c\to0$ limit of \eqref{eq:eshelby} recovers \eqref{eq:displacement-static-final}.

\subsubsection{Velocity, rotation and angular velocity fields}
The motion of each individual atom as the dislocation propagates through the lattice is described by a time varying $\tilde{\mathbf x}_i$, with velocity $\mathrm d\tilde{\mathbf x}_i/{\mathrm dt}=\mathrm d\mathbf u_i/{\mathrm dt}$. This allows to define the velocity field $\mathbf v$ such that $\mathrm d\tilde{\mathbf x}_i/{\mathrm dt}=\mathbf v(\mathbf x_i)$.

In this work, we intend to study the effective magnetic fields produced by the rotation of the lattice. Thus, we introduce the rotation field as \cite{landautheory1986}  $\boldsymbol\phi_R=\frac12\boldsymbol\nabla\times\mathbf u$, and the corresponding angular velocity $\boldsymbol\Omega_R=\dot{\boldsymbol\phi}_R=\frac12\boldsymbol\nabla\times\dot{\mathbf u}$. For the displacements \eqref{eq:eshelby}, this produces an instantaneous rotation $\boldsymbol\phi_R=\phi_R\,\hat{\mathbf z}$ and angular velocity $\boldsymbol\Omega_R=\Omega_R\,\hat{\mathbf z}$, with
\begin{equation}
	\phi_R=-\frac{a}{2\pi}\frac{c_t^2}{c^2}\frac{\tilde\gamma_t^2(1-\gamma_t^2)}{\gamma_t}
	\frac{x-ct}{(x-ct)^2+y^2\gamma_t^2}
\end{equation}
and
\begin{equation}
	\Omega_R=-\frac{a}{2\pi}\frac{c_t^2}{c}\frac{\tilde\gamma_t^2(1-\gamma_t^2)}{\gamma_t}
	\frac{(x-c t)^2-\gamma_t ^2 y^2}{\left((x-c t)^2+\gamma_t ^2 y^2\right)^2}.\label{eq:omega-full}
\end{equation}

Due to the time symmetry of the dislocation propagation throughout the lattice, we choose $t=0$ to study the structure of these fields, without loss of generality. The structure of the angular velocity around the dislocation core becomes more clear in polar coordinates $(r,\theta)$ such that $x=r\cos\theta$ and $y=r\sin\theta$. Defining, too, $\beta=c/c_t$, 
\begin{equation}
	\Omega_R=-\Omega_0\frac{a^2}{r^2}\beta\frac{  \left(1-\beta ^2/2\right) \left(\cos ^2\theta-\left(1-\beta ^2\right) \sin ^2\theta\right)}{\sqrt{1-\beta ^2} \left(1-(1- \cos2 \theta)\beta ^2/2\right)^2},\label{eq:omega-full-polar}
\end{equation}
with $\Omega_0=c_t/2\pi a$.

\begin{figure}\centering
	\includegraphics[width=\linewidth]{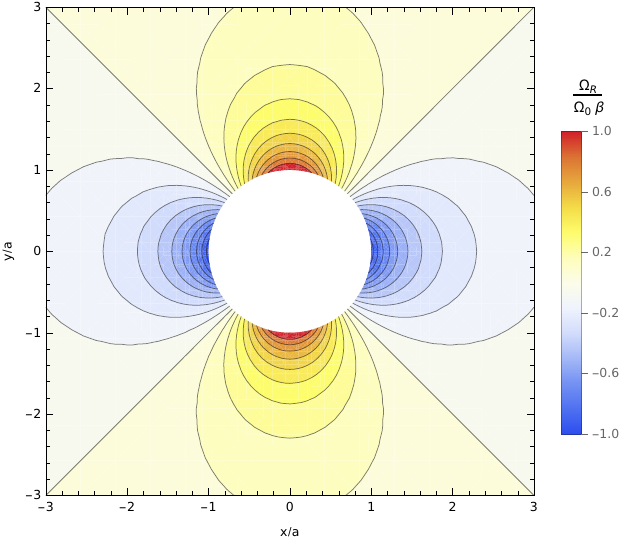}
	\caption{Scaled angular velocity $\Omega_R/\Omega_0\beta$ field around the singularity, for $r>a$.}
	\label{fig:ang-vel:1}
\end{figure}

In the low speed limit,
\begin{equation}
	\Omega_R=-\Omega_0\beta\frac{a^2}{r^2}\cos2\theta+O\left(\beta\right)^3.\label{eq:omega1-polar}
\end{equation}
This function has extrema at $\theta=0,\pi/2,\pi,3\pi/2$, and vanishes at $\theta=\pi/4,3\pi/4,5\pi/4,7\pi/4$. This produces a separation into two conical regions, within which the angular velocity presents a lobe shape, as shown in Fig.~\ref{fig:ang-vel:1} for a region with $r>a$ around the central singularity. We show the dimensionless quantity $\Omega_R/\beta\Omega_0$, which (at first order in $\beta$) is bounded between $-1$ and $1$.

Higher order terms in $\beta$ produce corrections that preserve, to some extent, the lobe structure, although \emph{(i)} the maximum angular velocity increases in magnitude in the central, clockwise region  ($\Omega_R>0$) faster than it does in the surrounding, counterclockwise regions ($\Omega_R<0$), and \emph{(ii)} the clockwise rotation region  becomes squeezed, and the counterclockwise rotation region  expands. In particular, these regions (in the upper half-plane---equivalent statements are valid for the lower half-plane) are separated by the angles such that $\Omega_R=0$, that is, at $\theta=\theta^*$ and $\theta=\pi-\theta^*$, where $\theta^*=\arctan\gamma_t$. In the low speed limit, $\theta^*\to\pi/4$, while in the high speed limit $\theta^*\to\pi/2$.

\subsubsection{The effective magnetic field}
Analysis of the Dirac equation in a non-inertial frame of reference (see \cite{hehlinertial1990}) leads to modifications in the Hamiltonian which, in the non-relativistic limit, couple the particle spin with the angular momentum as $\mathcal H_\mathrm{s-r}=-\hbar\, \mathbf S\cdot\mathbf\Omega_R$, in a manner similar to the usual coupling of the spin and an external magnetic field $\mathcal H_\mathrm{s-m}=-\hbar \gamma\,\mathbf S\cdot\mathbf H$.

Thus, the rotations produced by the moving dislocation may be studied as an effective magnetic field 
\begin{equation}
	\mathbf H_R=\frac{\boldsymbol\Omega_R}{\gamma},
\end{equation}
where $\gamma$ is the gyromagnetic ratio of the atoms in the lattice.
The spatial structure of this magnetic field is thus the same as that of $\mathbf\Omega$, described above. We now turn to assess the extent to which such effective magnetic fields are strong enough to reverse the magnetization of a sizable amount of atoms in the lattice.

As a simple approach to the problem, we may take the maximum (in magnitude) angular velocity in the low speed limit \eqref{eq:omega1-polar}, occurring at $r=a$ and $\theta=0$ to obtain the corresponding magnetic field. As an order of magnitude estimate, we take $c\sim c_t$ to obtain
\begin{equation}
	H_R\sim \frac{c_t}{2\pi a\gamma}. \label{HR}
\end{equation}
Using typical values ($c_t \sim 3 \times 10^3\,\unit{\meter\per\second}$, $a \sim 3 \,\si{\angstrom}$, and $\gamma =1.6 \times 10^{11}\,\unit{\per\second\per\tesla}$), $H_R\sim 10\,\mathrm T$. The inverse-square dependence of the angular velocity results in a sizeable effective magnetic field (above $0.1\mathrm T$) up to a distance of ten lattice spacings above or below the dislocation. Smaller $c_t$ and a greater size of the unit cell $a$ would result in a smaller effective field but it would still be large compared to a typical coercive field of a magnetic material. This provides a strong indication that dislocation propagation may produce strong enough effective magnetic fields to reverse the magnetization of solids.

\section{Spin reversal by a moving dislocation} \label{reversal}

\subsection{The Landau-Lifshitz equation}
\begin{figure}\centering
	\includegraphics[width=\linewidth]{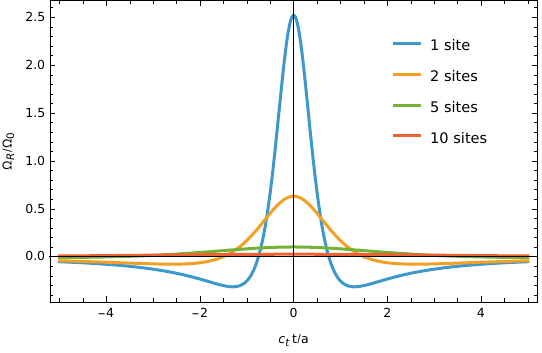}
	\caption{Time profile of the angular velocity for different atomic sites above/below the $y=0$ axis, for a dislocation propagating at $c=0.8c_t$.}
	\label{fig:time-profile}
\end{figure}

To study the effects of the moving  dislocation on individual spins, we must consider that the situation depicted in Fig.~\ref{fig:ang-vel:1} is only instantaneous. Any given lattice site will be subject to different angular velocities and effective magnetic fields as the dislocation passes through. 
For a rightward-moving dislocation, lattice sites will undergo three periods of rotation in opposite directions: clockwise--counterclockwise--clockwise. Due to the structure of the solutions described above, the initial and final clockwise motions will last longer but be slower, while the intermediate counterclockwise motion will be shorter but more intense. This behavior is depicted in Fig.~\ref{fig:time-profile}.

 The final magnetization state of an individual atom will thus depend in a complex manner on the evolution of the dislocation. While fast dislocations increase the effective field, slow dislocations may exert their influence for longer. In this section, we address those complexities via numerical analysis.

Consider an atomic spin ${\bf S}$ (with ${\bf S}^2 =1$) embedded in a non-magnetic matrix in which the dislocation is propagating. Its dynamics can be studied with the help of the Landau-Lifshitz equation \cite{Lectures} in the rotating frame rigidly coupled to the crystal lattice:
\begin{equation}
\frac{d {\bf S}}{d t}= {\bf S \times {\bm \Omega}} - {\alpha}{\bf S}{\times}\left({\bf S}{\times}{\bm \Omega}\right).\label{eq:LL}
\end{equation}
Here $\alpha$ is the phenomenological damping constant and ${\bf \Omega} = \gamma {\bf H}_{\rm eff}$  with the effective field ${\bf H}_{\rm eff}$ given by
\begin{equation}
{\bf H}_{\rm eff} = -\frac{\partial{\cal{H}}}{\partial{\bf S}},\label{eq:hamiltonian-heff}
\end{equation}
where ${{\cal{H}}}$ is the Hamiltonian in the lattice frame. 

In general, we may decompose the hamiltonian as $\mathcal H=\mathcal H_R+\mathcal H_H+\mathcal H_A$, where $\mathcal H_R$, $\mathcal H_H$ and $\mathcal H_A$ are the rotation, external field and anisotropy terms, respectively. Following the arguments above, the rotation term is $\mathcal H_R=-\hbar\,\mathbf S\cdot \boldsymbol\Omega_R$. The external field term, when present, is $\mathcal H_H=-\hbar\gamma\,\mathbf S\cdot \mathbf H$. We consider two different models for the magnetic anisotropy: a cubic anisotropy $\mathcal H_A^c=-\frac14D_c(S_x^4+S_y^4+S_z^4)$ and a uniaxial anisotropy along the $z$ axis, with $\mathcal H_A^z=-\frac12D_\shortparallel S_z^2$.

This calls for a similar decomposition $\boldsymbol\Omega=\boldsymbol\Omega_R+\boldsymbol\Omega_H+\boldsymbol\Omega_A$, where $\boldsymbol\Omega_R$ is obtained from \eqref{eq:omega-full} and $\boldsymbol\Omega_H=\gamma\mathbf H$. The two anisotropy models yield, using \eqref{eq:hamiltonian-heff}, $\boldsymbol\Omega_A^c=D_c(S_x^3\,\hat{\mathbf x}+S_y^3\,\hat{\mathbf y}+S_z^3\,\hat{\mathbf z})$ and $\boldsymbol\Omega_A^z=D_\shortparallel S_z\,\hat{\mathbf z}$.

Next, we turn to a numerical analysis of the solutions to \eqref{eq:LL} in two different cases: \emph{(i)} cubic anisotropy with $\mathbf H=\boldsymbol 0$, and \emph{(ii)} uniaxial anisotropy with $\mathbf H\neq\boldsymbol0$.

\subsection{Numerical results}
For numerical work, we use the dimensionless time $\tau= t\,c_t/a$, and define $\boldsymbol\omega=a\boldsymbol\Omega/c_t$ and $\boldsymbol\omega_k=a\boldsymbol\Omega_k/c_t$, for $k=A,H,R$. We also use the rescaled constants $\delta_k=aD_k/\hbar c_t$, with $k=c,\shortparallel$. With this, \eqref{eq:LL} turns into $\dot{\mathbf S}=\mathbf S\times\boldsymbol\omega-\alpha\,\mathbf S\times(\mathbf S\times\boldsymbol\omega)$, where $\cdot$ indicates $\tau$-derivatives. For illustration, we will study the evolution of a spin located one lattice space above the slip plane ($y=a$) and at $x=0$, and illustrate how the other parameters affect the occurrence of spin reversals.

\subsubsection{Cubic anisotropy}
We first consider a cubic anisotropy model without an external magnetic field. In this case, $\boldsymbol\omega=\delta_c(S_x^3\,\hat{\mathbf x}+S_y\,\hat{\mathbf y}+S_z\,\hat{\mathbf z})+\omega_R\,\hat{\mathbf z}$. Landau-Lifshitz's equation becomes
\begin{subequations}
	\begin{gather}
		\begin{split}
						\,\dot S_x=\ & \alpha  S_x \left\{\delta_c \left[S_x^2 \left(S_y^2+S_z^2\right)- S_y^4-S_z^4\right]-\omega_R S_z\right\}
							        \\& +\delta_c S_yS_z(S_z^2-S_y^2)+\omega _R S_y  ,
		\end{split}\\
		\begin{split}
						\,\dot S_y=\ & \alpha S_y  \left\{\delta_c \left[S_y^2 \left(S_z^2 +S_x^2\right) -S_z^4- S_x^4\right]-\omega_RS_z\right\}
									  \\& +\delta_c S_z S_x(S_x^2-S_z^2)-\omega_RS_x ,
		\end{split}\\
		\begin{split}
						\,\dot S_z=\ & \alpha S_z  \left\{\delta_c \left[S_z^2\left(S_x^2+S_y^2\right)- S_x^4-S_y^4\right]-\omega_RS_z\right\}
							        \\& +\delta_c S_x S_y( S_y^2- S_x^2)+\alpha\omega _R.
		\end{split}
	\end{gather}
\end{subequations}

This presents $26$ ($6+12+8$) possible steady ($\dot{\mathbf S}=\boldsymbol0$) states for $\mathbf S$ when $\omega_R=0$. These are of three types: $\left<1,0,0\right>$, $\left<1,1,0\right>/\sqrt2$ and $\left<1,1,1\right>/\sqrt3$. The set of all equilibrium states is spanned by all independent sign flips and permutations of these three. Nevertheless, only the first type is a stable steady state. Thus, we consider an initial spin $\mathbf S_0=\left<1,0,0\right>$ and study whether it can reach other of the steady states after the dislocation has passed.

\begin{figure}\centering
	\includegraphics[width=\linewidth]{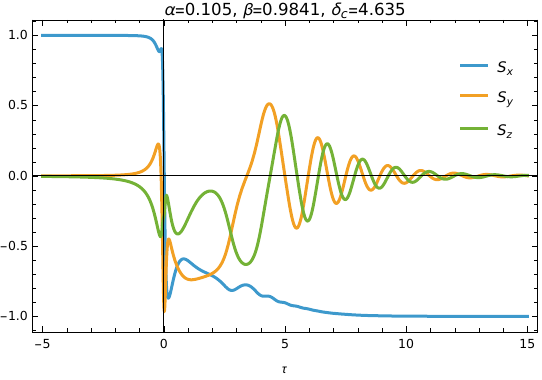}
	\caption{Time evolution of $\mathbf S$ for the cubic anisotropy model for $(\alpha,\beta,\delta_c)=(0.105,0.9841,4.635)$, illustrating a reversal from $S_x=1$ to $S_x=-1$.}
	\label{fig:cubic:reversal}
\end{figure}

We explore the parameter space for $0<\alpha<10$, $0<\beta<1$ and $0<\delta_c<10$. 
Full reversals $\left<1,0,0\right>\to\left<-1,0,0\right>$ are observed to occur for $\beta\gtrsim 0.92$. The slowest among these occur for $\alpha\sim10^{-4}-10^{-3}$ and $\delta_c\sim3-5$. Larger values of $\alpha$ or smaller values of $\delta_c$ require much larger propagation speeds. For illustration, we choose $(\alpha,\beta,\delta_c)=(0.105,0.9841,4.635)$ and present the evolution of the spin in Fig.~\ref{fig:cubic:reversal}.

\begin{figure}\centering
	\includegraphics[width=\linewidth]{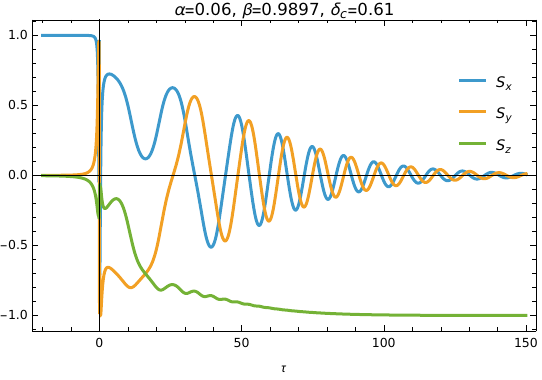}
	\caption{Time evolution of $\mathbf S$ for the cubic anisotropy model for $(\alpha,\beta,\delta_c)=(0.06,0.9897,0.61)$, illustrating a rotation from $S_x=1$ to $S_z=1$.}
	\label{fig:cubic:rotation}
\end{figure}

Rotations from the $x$ to the $y$ or $z$ directions occur as well throughout the $(\alpha,\beta,\delta_c)$ parameter space. In particular, $\left<1,0,0\right>\to\left<0,\pm1,0\right>$ and $\left<1,0,0\right>\to\left<0,0,-1\right>$ rotations occur in a similar range of parameters as the full reversals along the $x$ axis. Rotations $\left<1,0,0\right>\to\left<0,0,1\right>$, on the other hand, are observed at lower speeds as well, for sufficiently large $\alpha$. An illustration of this phenomenon, albeit for lower $\alpha$ and larger speed, is shown in Fig.~\ref{fig:cubic:rotation} for $(\alpha,\beta,\delta_c)=(0.06,0.9897,0.61)$. 

\subsubsection{Uniaxial $z$-anisotropy with external field}
We now study the uniaxial anisotropy model subject to an external magnetic field. 

Here, we pause to address a subtlety not relevant for our prior discussions: the evolution of the spin is dictated by the Landau-Lifshitz equation derived via \eqref{eq:hamiltonian-heff} from a hamiltonian in the reference frame of the lattice. The external magnetic field $\mathbf H$ is a constant in the laboratory frame. Thus, it will vary in direction if studied in the lattice frame as the dislocation passes, making the direction of the term $\boldsymbol\omega_H$ time dependent.

Taking $\{\hat{\mathbf x},\hat{\mathbf y},\hat{\mathbf z}\}$ and $\{\hat{\mathbf x}',\hat{\mathbf y}',\hat{\mathbf z}'\}$ to be the local basis vectors in the lattice and laboratory frames, respectively, we choose an external magnetic field along the laboratory $y$ direction: $\mathbf H=H\hat{\mathbf y}'$. The angle $\phi_R$ measures the counterclockwise rotation of the lattice frame with respect to the laboratory frame, so $\hat{\mathbf y}'= \sin\phi_R\,\hat{\mathbf x}+\cos\phi_R\,\hat{\mathbf y}$. With this, $\boldsymbol\omega=(\omega_R+\delta_\shortparallel S_z)\,\hat{\mathbf z}+\omega_H(\sin\phi_R\,\hat{\mathbf x}+\cos\phi_R\,\hat{\mathbf y})$.

The Landau-Lifshitz equation for this system is
\begin{subequations}
	\begin{gather}
		\begin{aligned}
							\dot S_x=\ &\omega_H \left[\alpha  \left(S_y^2+S_z^2\right)\sin\phi_R-\left(\alpha  S_x S_y+S_z\right)\cos\phi_R  \right] 
										\\&-\left(\alpha  S_x S_z-S_y\right) \left(\delta_\shortparallel  S_z+\omega_R\right),
		\end{aligned}\\
		\begin{aligned}
							\dot S_y=\ &\omega_H \left[\alpha  \left(S_x^2+S_z^2\right)\cos\phi_R-\left(\alpha  S_x S_y-S_z\right)\sin\phi_R  \right]
										\\&-\left(\alpha  S_y S_z+S_x\right) \left(\delta_\shortparallel  S_z+\omega_R\right) ,
		\end{aligned}\\
		\begin{aligned}
							\ \dot S_z=\ &\omega_H \left[\left(S_x-\alpha  S_y S_z\right)\cos\phi_R-\left(\alpha  S_x S_z+S_y\right)\sin\phi_R  \right]
										\\&+\alpha  \left(S_x^2+S_y^2\right) \left(\delta_\shortparallel  S_z+\omega_R\right).
		\end{aligned}
	\end{gather}
\end{subequations}

\begin{figure}[htbp!]
	\center
	\includegraphics[width=\linewidth]{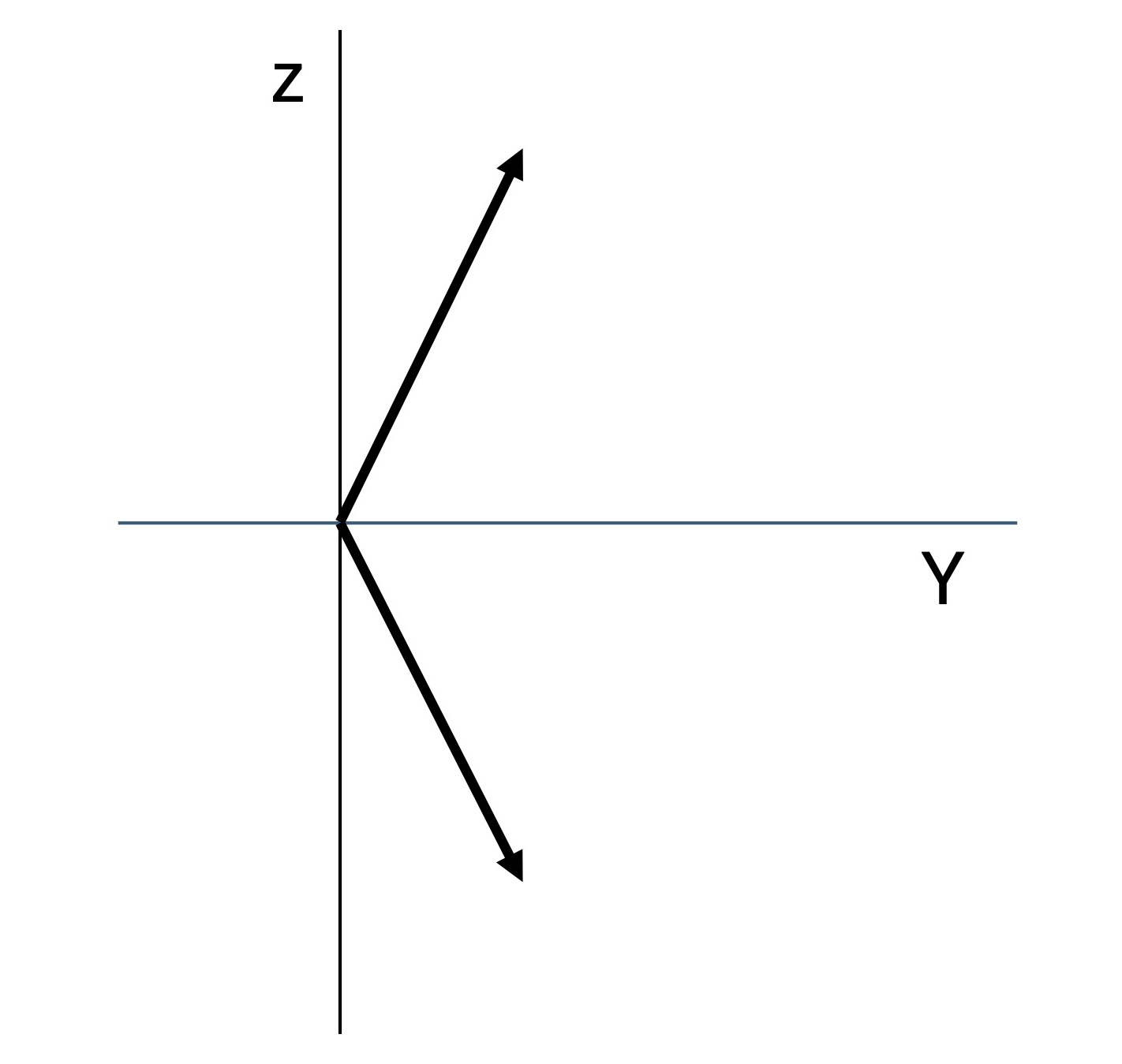}
	\caption{Degenerate equilibrium spin orientations in the presence of the field in the Y-direction.}
	\label{CantedS}
\end{figure}

This presents four steady states: $\left<0,\pm1,0\right>$ and
\begin{equation}
	\left<0,\frac{\omega_H}{\delta_\shortparallel},\pm\sqrt{1-\frac{\omega_H^2}{\delta_\shortparallel^2}}\right>. 
\end{equation}
The latter are only possible when $|\omega_H|<\delta_\shortparallel$. At larger magnetic fields, the degeneracy in $S_z$ disappears, as illustrated in Fig.~\ref{CantedS}, and the energy minimum is achieved by $\mathbf S$ pointing in the $y$-direction.

We will consider initial conditions with $S_z>0$, an study the extent to which reversals may occur for as large values of $S_z$ as possible (that is, for the smallest possible $|\omega_H|/\delta_\shortparallel$).
 
\begin{figure}\centering
	\includegraphics[width=\linewidth]{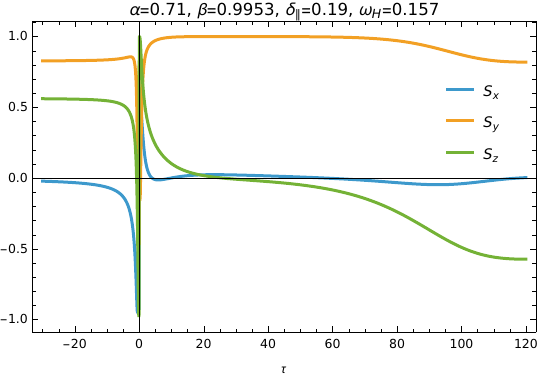}
	\caption{Time evolution of $\mathbf S$ for the uniaxial anisotropy model with external magnetic field for $(\alpha,\beta,\delta_\shortparallel,\omega_H)=(0.71,0.9953,0.19,0.157)$, illustrating a rotation from $S_z\approx0.56$ to $S_z\approx-0.56$.}
	\label{fig:extH:reversal}
\end{figure}

\begin{figure}\centering
	\includegraphics[width=\linewidth]{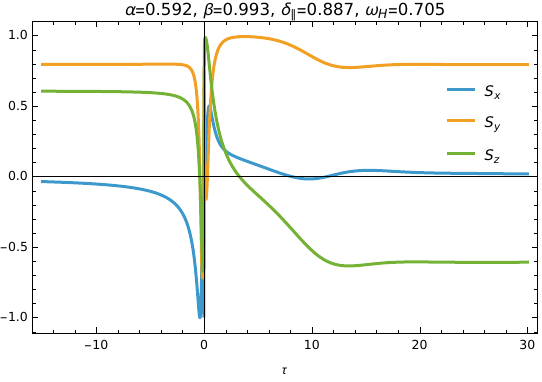}
	\caption{Time evolution of $\mathbf S$ for the uniaxial anisotropy model with external magnetic field for $(\alpha,\beta,\delta_\shortparallel,\omega_H)=(0.592,0.993,0.887,0.705)$, illustrating a rotation from $S_z\approx0.61$ to $S_z\approx-0.61$.}
	\label{fig:extH:reversal-fast}
\end{figure}

Like before, we study the occurrence of reversals throughout the $(\alpha,\beta,\delta_\shortparallel,\omega_H)$ parameter space. We recall that the steady states for $|\omega_H|<\delta_\shortparallel$ correspond to $S_z=\pm\sqrt{1-{\omega_H^2}/{\delta_\shortparallel^2}}$. It is reasonable to expect reversals to occur with ease at low speeds if $|\omega_H|$ is close enough to $\delta_\shortparallel$. Conversely, regardless of how small $\omega_H$ is (that is, of how large $S_z$ is initially), spins are reversible at large enough speeds. We illustrate this effect with in Figs.~\ref{fig:extH:reversal} and \ref{fig:extH:reversal-fast} for two sets of parameters $(\alpha,\beta,\delta_\shortparallel,\omega_H)=(0.71,0.9953,0.19,0.157)$ and $(0.592,0.993,0.887,0.705)$, respectively. Both situations show the reversal of $S_z\sim0.6$ to $S_z\sim-0.6$; the different parameter choices show a difference in about an order of magnitude in the reversal time.

\section{Discussion} \label{discussion}

We have shown that fast-moving edge dislocations generate large effective magnetic fields in the tesla range as effectively as the high-frequency chiral phonons \cite{Davies-Nature2024,Juraschek-PRR2022,Luo-Science2023}. This is true for all solids and is based on Eq.\ (\ref{HR}) that has high generality as has been explained via a simple argument in the introduction. 

Massive formation of dislocations moving at speeds close to the speed of sound occurs when a solid experiences an elastic stress approaching the limit of plastic deformation. It can be induced by a voltage, laser, microwave, or terahertz acoustic pulse. Giant effective magnetic fields generated by moving dislocations are typically greater than the fields required to reverse individual spins or a magnetization of a thin ferromagnetic layer. 

The reversal of individual spins has been demonstrated by solving the dynamical equation for the spin in the coordinate frame rigidly coupled to the crystallographic axes. In such a non-inertial frame, the spin relaxes to the direction of the total effective field modified by the elastic twist from a passing dislocation. Such effect of the rotation is the essence of the Barnet effect. The characteristic time of the spin reversal is $a/(c_t\alpha)$, with $a$ being the size of the unit cell of the crystal, $c_t$ being the speed of the transverse sound, and $\alpha$ being the damping parameter. 

At, e.g.,  $a \sim 3 \,\si{\angstrom}$, $c_t \sim 3 \times 10^3\,\unit{\meter\per\second}$, and $\alpha \sim 0.1$ the reversal time is in the picosecond range. It is, therefore, plausible that the ultrafast magnetization reversal observed in experiments that generated strong elastic shocks has been generated by massive formation of fast-moving dislocations. A full-scale simulation of such dynamics is rather involved and will be done elsewhere.

\section{Acknowledgments}

The authors are grateful to Dmitry Garanin for helpful discussion. This work has been supported by Grant No. FA9550-24-1-0290 funded by the Air Force Office of Scientific Research.

\end{document}